\documentclass[smus]{snow2e}
\usepackage{graphics}

\newcommand{\sss}{\scriptscriptstyle}
\def\fH{{\cal H}}

\newcommand{\be}{\begin{equation}}
\newcommand{\ee}{\end{equation}}
\newcommand{\bea}{\begin{eqnarray}}
\newcommand{\eea}{\end{eqnarray}}
\newcommand{\ba}{\begin{array}}
\newcommand{\ea}{\end{array}}

\newcommand{\pl}{Phys.\ Lett.}
\newcommand{\np}{Nucl.\ Phys.}
\newcommand{\prl}{Phys.\ Rev.\ Lett.}
\newcommand{\prd}{Phys.\ Rev.\ D}

\newcommand{\etal}{{\it et al}.,\ }

\begin{document}

\title{Signals from Flavor Changing Scalar Currents at the Future 
Colliders\thanks{Work supported by U.S. Department of
Energy contracts DC-AC05-84ER40150 (CEBAF) and DE-AC-76CH0016 (BNL).}}
\author{D. Atwood\\{\it Theory Group, Continuous Electron Beam
Accelerator Facility, Newport News, VA 23606}\\ L. Reina and A. Soni\\
{\it Physics Department, Brookhaven National Laboratory, Upton, NY
11973}}

\maketitle
\thispagestyle{empty}\pagestyle{empty}

\begin{abstract} 
We present a general phenomenological analysis of a class of Two Higgs
Doublet Models with Flavor Changing Neutral Currents arising at the
tree level. The existing constraints mainly affect the couplings of
the first two generations of quarks, leaving the possibility for non
negligible Flavor Changing couplings of the top quark open. The next
generation of lepton and hadron colliders will offer the right
environment to study the physics of the top quark and to unravel the
presence of new physics beyond the Standar Model. In this context we
discuss some interesting signals from Flavor Changing Scalar Neutral
Currents.
\end{abstract}

\section{General Framework}

The next generation of lepton and hadron colliders will play a
fundamental role in the study of new physics beyond the Standar Model
(SM). Higher energies will allow a careful study of the physics of the
top quark (its couplings in particular) and of the scalar and gauge
sector of the fundamental theory of elementary particles.

In this context, we have analyzed the possibility of having a Two
Higgs Doublet Model (2HDM) with Flavor Changing Neutral Currents
(FCNC's) allowed at the tree level \cite{sher}-\cite{hall}. This Model
constitutes a simple extension of the scalar sector of the Standard
Model and closely mimics the Higgs sector of a SuperSymmetric Theory
(SUSY). However, the possibility of having flavor changing (FC) tree
level couplings in the neutral scalar sector definitely distinguishes
it from both the SM and SUSY. Moreover, the discovery and study of
extra scalar or pseudoscalar, neutral and charged particles with not
too heavy masses will be in the reach of the future machines. {}From
here our interest.

Although there is no {\it a priori} veto to the existence of FCNC at
the tree level, the low energy phenomenology of the K- and of the
B-meson as well as the existing precision measurements of the SM
impose strong constraints on the possibility of having sizable effects
from FCNC. However, under suitable assumptions, the FC couplings of
the top quark partially escape these constraints and can be predicted
to give non negligible signals as we will illustrate in the following.

\subsection{The Model}

A mild extension of the SM with one additional scalar SU(2) doublet
opens up the possibility of flavor changing scalar currents (FCSC's)
at the tree level. In fact, when the up-type quarks and the down-type
quarks are allowed simultaneously to couple to more than one scalar
doublet, the diagonalization of the up-type and down-type mass
matrices does not automatically ensure the diagonalization of the
couplings with each single scalar doublet. For this reason, the 2HDM
scalar potential and Yukawa Lagrangian are usually constrained by an
{\it ad hoc} discrete symmetry \cite{glash}, whose only role is to
protect the model from FCSC's at the tree level. Let us consider a
Yukawa Lagrangian of the form

\bea
\label{lyukmod3}
{\cal L}_{Y}&=& \eta^{U}_{ij} \bar Q_{i,L} \tilde\phi_1 U_{j,R} +
\eta^D_{ij} \bar Q_{i,L}\phi_1 D_{j,R} + \\ 
&& \xi^{U}_{ij} \bar Q_{i,L}\tilde\phi_2 U_{j,R}
+\xi^D_{ij}\bar Q_{i,L} \phi_2 D_{j,R} \,+\, h.c. \nonumber
\eea

\noindent where $\phi_i$, for $i=1,2$, are the two scalar doublets of
a 2HDM, while $\eta^{U,D}_{ij}$ and $\xi_{ij}^{U,D}$ are the non
diagonal matrices of the Yukawa couplings.  Imposing the following
{\it ad hoc} discrete symmetry

\bea
\phi_1 \rightarrow -\phi_1 \,\,\,\,\,\,\,\,\,\mbox{and}&&
\,\,\,\,\,\phi_2\rightarrow\phi_2\\ \label{discr_symm}
D_i\rightarrow -D_i \,\,\,\,\,\,\,\,\,\mbox{and}&&\,\,\,\,\, 
U_i\rightarrow\mp U_i\nonumber
\eea

\noindent some of the terms in ${\cal L}_{Y}$ have to be dropped and
one obtains the so called Model~I and Model~II, depending on whether
the up-type and down-type quarks are coupled to the same or to two
different scalar doublets respectively \cite{hunter}.

In contrast we will consider the case in which no discrete symmetry 
is imposed and both up-type and down-type quarks then have FC
couplings. For this type of 2HDM, which we will call Model~III, the
Yukawa Lagrangian for the quark fields is as in Eq.~(\ref{lyukmod3})
and no term can be dropped {\it a priori}, see also
refs.~\cite{lukesavage,eetc}~.

For convenience we can choose to express $\phi_1$ and $\phi_2$ in a
suitable basis such that only the $\eta_{ij}^{U,D}$ couplings generate
the fermion masses, i.e.\ such that

\be
\langle\phi_1\rangle=\left( 
\begin{array}[]{c}
0\\
{v/\sqrt{2}}
\end{array}
\right)\,\,\,\, , \,\,\,\,
\langle\phi_2\rangle=0 \,\,\,.
\ee

\noindent The two doublets are in this case of the form

\bea
\phi_1&=&\frac{1}{\sqrt{2}}\left[\left(\ba{c} 0 \\ v+H^0\ea\right)+
\left(\ba{c} \sqrt{2}\,\chi^+\\ i\chi^0\ea\right)\right]\nonumber\\
\phi_2&=&\frac{1}{\sqrt{2}}\left(\ba{c}\sqrt{2}\,H^+\\ H^1+i H^2\ea
\right)\,\,\,.
\eea

\noindent The scalar Lagrangian in the ($H^0$, $H^1$, $H^2$,
$H^{\pm}$) basis is such that \cite{knowles,hunter}~: the doublet
$\phi_1$ corresponds to the scalar doublet of the SM and $H^0$ to the
SM Higgs field (same couplings and no interactions with $H^1$ and
$H^2$); all the new scalar fields belong to the $\phi_2$ doublet; both
$H^1$ and $H^2$ do not have couplings to the gauge bosons of the form
$H^{1,2}ZZ$ or $H^{1,2}W^+W^-$.

\noindent $H^{\pm}$ is the charged scalar mass eigenstate, while the
two scalar plus one pseudoscalar neutral mass eigenstates are obtained
from ($H^0$, $H^1$, $H^2$) as follows

\bea 
\label{masseigen}
\bar H^0 & = & \left[(H^0-v)\cos\alpha + H^1\sin\alpha \right]
\nonumber \\ 
h^0 & = & \left[-(H^0-v)\sin\alpha + H^1\cos\alpha \right] \\ 
A^0 & = &  H^2 \nonumber 
\eea 

\noindent where $\alpha$ is a mixing angle, such that for
$\alpha\!=\!0$, ($H^0$, $H^1$, $H^2$) coincide with the mass
eigenstates.

Furthermore, to the extent that the definition of the $\xi^{U,D}_{ij}$
couplings is arbitrary, we will denote by $\xi^{U,D}_{ij}$ the new
rotated couplings, such that the charged couplings look like
$\xi^{U}\cdot V_{\sss{\rm CKM}}$ and $V_{\sss{\rm CKM}}\cdot\xi^{D}$.
This form of the charged couplings is indeed peculiar to Model~III
compared to Models~I and II and can have important phenomenological
repercussions \cite{rbrc,longwu}.

In order to apply to specific processes we have to make some definite
ansatz on the $\xi_{ij}^{U,D}$ couplings. Many different suggestions
can be found in the literature \cite{sher,antaramian,hall,eetc}. In
addition to symmetry arguments, there are also arguments based on the
widespread perception that these new FC couplings are likely to mainly
affect the physics of the third generation of quarks only, in order to
be consistent with the constraints coming from $K^0\!-\!\bar K^0$ and
$B^0_d\!-\!\bar B^0_d$. A natural hierarchy among the different quarks
is provided by their mass parameters, and that has led to the
assumption that the new FC couplings are proportional to the mass of
the quarks involved in the coupling. Most of these proposals are well
described by the following ansatz

\be
\xi^{U,D}_{ij}=\lambda_{ij}\,\frac{\sqrt{m_i m_j}}{v} 
\label{coupl_sher}
\ee 

\noindent which basically coincides with what was proposed by Cheng
and Sher \cite{sher}. In this ansatz the residual degree of
arbitrariness of the FC couplings is expressed through the
$\lambda_{ij}$ parameters, which need to be constrained by the
available phenomenology. In particular we will see how $K^0\!-\!\bar
K^0$ and $B^0_d\!-\!\bar B^0_d$ mixings (and to a less extent
$D^0\!-\!\bar D^0$ mixing) put severe constraints on the FC couplings
involving the first family of quarks. Additional constraints are given
by the combined analysis of the $Br(B\rightarrow X_s\gamma)$, the
$\rho$ parameter, and $R_b$, the ratio of the $Z\rightarrow b\bar b$
rate to the $Z$ hadronic rate. We will analyze all these constraints
in the following section

\subsection{Discussion of the Constraints}

The existence of FC couplings is very much constrained by the
experimental results on $F^0\!-\!\bar F^0$ flavor mixings (for
$F\!=\!K,B$ and to a less extent $D$)

\bea
\label{mixing_exp}
\Delta M_K&\simeq& 3.51\cdot 10^{-15}\,\,\mbox{GeV}\nonumber\\
\Delta M_{B_d}&\simeq& 3.26\cdot 10^{-13}\,\,\mbox{GeV}\\ 
\Delta M_D&<& 1.32\cdot 10^{-13}\,\,\mbox{GeV}\nonumber
\eea

\noindent due to the presence of new tree level contributions to each
of the previous mixings. We have analyzed the problem in detail
\cite{longwu}, taking into account both tree level and loop
contributions. Indeed the two classes of contributions can affect
different FC couplings, due to the peculiar structure of the charged
scalar couplings (see previous section).

We find that, unless for scalar masses in the multi-TeV range, the
tree level contributions need to be strongly suppressed, requiring
that the corresponding FC couplings are much less than one. Enforcing
the ansatz made in Eq.~(\ref{coupl_sher}), this amounts to demand that

\be
\lambda^D_{ds}\ll 1\,\,\,\,,\,\,\,\,\lambda^D_{db}\ll 1\,\,\,\,
\mbox{and}\,\,\,\, \lambda^U_{ud}\ll 1\,\,.
\label{coupl_fg}
\ee

\noindent More generally, we can assume that the FC couplings
involving the first generation are negligible. Particular 2HDM's have
been proposed in the literature in which this pattern can be realized
\cite{kao}. The remaining FC couplings, namely $\xi^U_{ct}$ and
$\xi^D_{sb}$ are not so drastically affected by the $F^0\!-\!\bar F^0$
mixing phenomenology.  {}From the analysis of the loop contributions to
the $F^0\!-\!\bar F^0$ mixings (box and penguin diagrams involving the
new scalar fields) we verify that many regions of the parameter space
are compatible with the results in Eq.~(\ref{mixing_exp})
\cite{longwu}. Therefore we may want to look at other constraints in
order to single out the most interesting scenarios.

Three are in particular the physical observables that impose strong
bounds on the masses and couplings of Model III \cite{rbrc,longwu}
\begin{itemize}
\item The inclusive branching ratio for $B\rightarrow X_s\gamma$,
which is measured to be \cite{alam}

\be 
Br(B\rightarrow X_s\gamma)=(2.32\pm 0.51\pm 0.29\pm 0.32)\times
10^{-4}
\label{bsg_exp}
\ee

\item The ratio

\be
R_b=\frac{\Gamma(Z\rightarrow b\bar b)}{\Gamma(Z\rightarrow
\mbox{hadrons})}
\ee

whose present measurement \cite{warsaw} is such that $R_b^{\rm
expt}>R_b^{\rm SM}$ ($\sim 1.8\sigma$)\footnote{ The value of
$R_b^{\rm expt}$ reported in Eq.~(\ref{rb_expSM}) corresponds to the
experimental measurement obtained for $R_c=R_c^{\rm SM}=0.1724$.}

\bea
\label{rb_expSM}
R_b^{\rm expt}&=& 0.2178\pm 0.0011 \\
R_b^{\rm SM}&=&0.2156\pm 0.0002 \nonumber
\eea

\noindent The value of $R_b^{\rm expt}$ seems to challenge many
extensions of the SM \cite{bamert,rbrc}. However, several issues on
the measurement of this observable are still unclear and require
further scrutiny \cite{rbrc}.

\item The corrections to the $\rho$ parameter. In fact, the
relation between $M_W$ and $M_Z$ is modified by the presence of new
physics and the deviation from the SM prediction is usually described
by introducing the parameter $\rho_0$ \cite{pdg,lang}, defined as

\be
\rho_0=\frac{M_W^2}{\rho M_Z^2\cos^2\theta_W} \label{rhozero_def}
\ee

\noindent where the $\rho$ parameter absorbs all the SM corrections
to the gauge boson self energies. In the presence of new physics

\be
\rho_0=1+\Delta\rho_0^{\sss{\rm NEW}} \label{rhozero}
\ee

\noindent {}From the recent global fits of the electroweak data, which
include the input for $m_t$ from Ref.~\cite{cdf} and the new
experimental results on $R_b$, $\rho_0$ turns out to be very close to
unity \cite{lang,rbrc,longwu}. This impose severe constraints on many
extension of the SM, especially on the mass range of the new
particles.
\end{itemize}
\noindent As is the case in 2HDM's with no FCNC's, it is very
difficult to reconcile the measured values of the previous three
observables in the presence of an extended scalar sector.  Taking into
account also the constraints from the $F^0\!-\!\bar F^0$ mixings, two
main scenarios emerge depending on the choice of enforcing or not
$R_b^{\rm exp}$ \cite{longwu}.

\begin{itemize}
\item [{\bf 1.}] If we {\bf enforce the constraint from $R_b^{\rm
expt}$} (see Eq.~(\ref{rb_expSM})), then we can accommodate the
present measurement of the $Br(B\rightarrow X_s\gamma)$ (see
Eq.~(\ref{bsg_exp})) and of the $\Delta F\!=\!2$ mixings (see
Eq.~(\ref{mixing_exp})) and at the same time satisfy the global fit
result for the $\rho$ parameter \cite{lang} provided the following
conditions are satisfied.
\begin{enumerate}
\item[i)] The neutral scalar $h^0$ and the pseudoscalar $A^0$ are very
light, i.e.

\be
50\, \mbox{GeV} \le M_h\sim M_A < 70 \,\mbox{GeV}\,\,.
\label{neutmass}
\ee

\item[ii)] The charged scalar $H^+$ is heavier than $h^0$ and $A^0$,
but not too heavy to be in conflict with the constraints from the
$\rho$ parameter. Thus

\be
150\, \mbox{GeV}\le m_c\le 200 \,\mbox{GeV} \,\,.
\label{chargmass}
\ee

\item[iii)] The $\xi^D_{ij}$ couplings are enhanced with respect to
the $\xi^U_{ij}$ ones

\bea 
\label{rbscenario}
\lambda_{bb}&\gg& 1\,\,\,\, \mbox{and}\,\,\,\, \lambda_{tt}\ll 1\\ 
\lambda_{sb}&\gg& 1\,\,\,\, \mbox{and}\,\,\,\, \lambda_{ct}\ll 1\,\,.
\nonumber 
\eea 
\end{enumerate}

\noindent The choice of the phase $\alpha$ is not as crucial as the
above conditions and therefore we do not make any assumption on it.

\item [{\bf 2.}]If we {\bf disregard the constraint from $R_b^{\rm
expt}$} there is no need to impose the bounds of
Eqs. (\ref{neutmass})-(\ref{rbscenario}) and we can safely work in the
scenario in which only the first generation FC couplings are
suppressed

\be
\lambda_{ui},\lambda_{dj}\ll 1\quad\mbox{for}\quad i,j=1,2,3
\ee

\noindent in order to satisfy the experimental constraints on the
$F^0\!-\!\bar F^0$ mixings. We will assume the FC couplings of the
second an third generations to be given by Eq.~(\ref{coupl_sher}) with

\be
\lambda_{ct}\simeq O(1)\,\,\,\,\mbox{and}\,\,\,\,
\lambda_{sb}\simeq O(1)\,\,. 
\label{ctsb_2}
\ee

\noindent The value of the mixing angle $\alpha$ is not relevant,
while the masses are mainly dictated by the fit to $Br(B\rightarrow
X_s\gamma)$ and $\Delta\rho_0$ \cite{rbrc}

\be
M_H,M_h\le M_c\le M_A\,\,\,\,\,\mbox{and}\,\,\,\,
M_A\le M_c\le M_H,M_h \,\,\,.
\label{Mc_inbetween}
\ee

\end{itemize}

\noindent We can see that, except in a very narrow window of the
parameter space, it is in general very difficult to accomodate the
present value of $R_b^{\rm exp}$ in Model III. Due to the present
unclear experimental situation for $R_b$, we will mainly concentrate
on the second scenario\footnote{See ref. \cite{longwu} for a
discussion of the scenario which accomodates $R_b^{\rm exp}$.}. This
scenario has the very interesting characteristics of providing sizable
FC couplings for the top quark, in a way that will certainly be
testable at the next generation of lepton and hadron colliders. We
will discuss some of these phenomenological issues in the next
section.

\section{Signals of Top-Charm Production}

If we assume $\lambda_{sb}\!\simeq\!O(1)$ and
$\lambda_{ct}\!\simeq\!O(1)$ as in Eq.~(\ref{ctsb_2}), $\xi^U_{ct}$
becomes the most relevant FC coupling. The presence of a $\xi^U_{ct}$
flavor changing coupling can be tested by looking at both top decays
and top production (see ref. \cite{longwu} and references therein).
We want to concentrate here on top-charm production at lepton
colliders, both $e^+e^-$ and $\mu^+\mu^-$, because, as we have
emphasized before \cite{eetc,mumutc}, in this environment the
top-charm production has a particularly clean and distinctive
signature. The SM prediction for this process is extremely suppressed
and any signal would be a clear evidence of new physics with large FC
couplings in the third family. Moreover it has a very distinctive
signature, with a very massive jet recoiling against an almost
massless one (very different from a $bs$ signal, for instance). This
characteristic is enhanced even more in the experimental environment
of a lepton collider.

In principle, the production of top-charm pairs arises both at the
tree level, via the $s$ channel exchange of a scalar field with FC
couplings, and at the one loop level, via corrections to the $Ztc$ and
$\gamma tc$ vertices.  The $s$ channel top-charm production is one of
the new interesting possibilities offered by a $\mu^+\mu^-$ collider
in studying the physics of standard and non standard scalar
fields. However, it is not relevant for an $e^+e^-$ collider, because
the coupling of the scalar fields to the electron is likely to be very
suppressed (see Eq.~(\ref{coupl_sher})). Therefore we will consider
these two cases separately.
\vspace{.5cm}

In the case of an $e^+e^-$ collider, top-charm production arises via
$\gamma$ and $Z$ boson exchange, i.e.\ the process
$e^+e^-\rightarrow\gamma^*,Z^*\rightarrow \bar t c+\bar c t$, where
the effective one loop $\gamma tc$ or $Ztc$ vertices are induced by
scalars with FC couplings. We will consider the total cross section
normalized to the cross section for producing $\mu^+\mu^-$ pairs via
one photon exchange, i.e.

\be R^{tc} \equiv \frac{\sigma(e^+e^-\rightarrow t\bar c+ \bar
tc)}{\sigma( e^+e^- \rightarrow \gamma^*\rightarrow \mu^+\mu^-)}
\label{rtc_ee} \ee

\noindent and normalized to $\lambda_{ij}\simeq\lambda\!=\!1$ (see
Eq.~(\ref{coupl_sher})), consistently with our Eq.~(\ref{ctsb_2}). For
the moment, we want to simplify our discussion by taking the same
$\lambda$ for all of the $\xi^{U,D}_{ij}$ couplings. Moreover, we want
to factor out this parameter, because it summarizes the degree of
arbitrariness we have on these new couplings and it will be useful for
further discussion.
\begin{figure}[htb]
\leavevmode
\begin{center}
\resizebox{!}{6cm}{%
\includegraphics{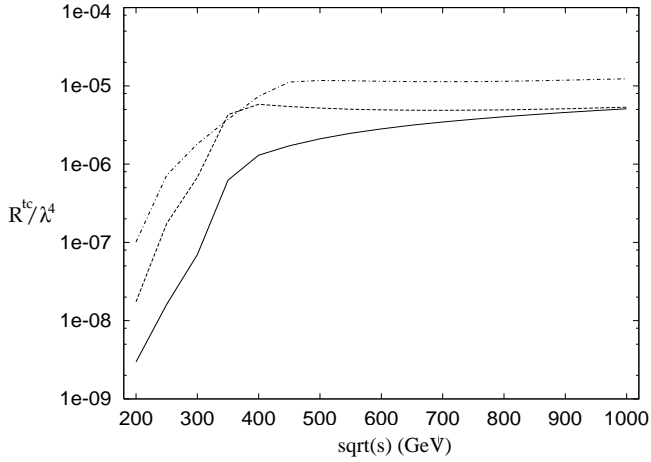}}
\caption[]{ $R_{tc}/\lambda^4$ vs.\ $\sqrt{s}$ when $M_h\!=\!200$ GeV
and $M_A\!\simeq\!M_c\!=\!1$ TeV (solid), $M_A\!=\!200$ GeV and
$M_h\!\simeq\!M_c\!=\!1$ TeV (dashed), $M_c\!=\!200$ GeV and
$M_h\!\simeq\!M_A\!=\!1$ TeV (dot-dashed).}
\label{eetc_plot}
\end{center}
\end{figure}

As already discussed in Ref.~\cite{eetc}, we take $m_t\!\simeq\!180$
GeV and vary the masses of the scalar and pseudoscalar fields in a
range between 200 GeV and 1 TeV\null. Larger values of the scalar
masses are excluded by the requirement of a weak coupled scalar
sector. The phase $\alpha$ does not play a relevant role and in our
qualitative analysis we will set $\alpha\!=\!0$. In
Fig.~\ref{eetc_plot} we plot $R^{tc}/\lambda^4$ as a function of
$\sqrt{s}$ for a sample of relevant cases, in which one of the scalar
particles is taken to be light ($M_l\!\simeq\! 200$ GeV) compared to
the other two ($M_h\!\simeq\!1$ TeV). We find that even with different
choices of $M_h$, $M_A$ and $M_c$ it is difficult to push
$R^{tc}/\lambda^4$ much higher than $10^{-5}$. Therefore the three
cases illustrated in Fig.~\ref{eetc_plot} appear to be a good sample
to illustrate the type of predictions we can obtain for the rate for
top-charm production in model~III\null.

{}From Fig.~\ref{eetc_plot}, we also see that going to energies much
larger than $\sim 400$--500 GeV (i.e.\ $\sim2M_l$) does not gain much
in the rate and in this case $R^{tc}/\lambda^4$ can be as much as
$10^{-5}$. Since it is reasonable to expect $10^4$--$10^5$
$\mu^+\mu^-$ events in a year of running for the next generation of
$e^+e^-$ colliders ($\int{\cal L}\simeq 5\times
10^{33}\,\mbox{cm}^{-2}\mbox{sec}^{-1}$) at $\sqrt{s}=500$ GeV, this
signal could be at the detectable level only for not too small values
of the arbitrary parameter $\lambda$.  Thus we can expect experiments
to be able to constrain $\lambda\le 1$, for scalar masses of a few
hundred GeVs.

\vspace{.5cm} Another interesting possibility to study top-charm
production is offered by Muon Colliders \cite{mumutc}. Although very
much in the notion stage at present, $\mu^+\mu^-$ colliders has been
suggested as a possible lepton collider for energies in the TeV range
\cite{palmer,barger}.  Most of the applications of Muon Colliders
would be very similar to electron colliders. One advantage, however,
is that they may be able to produce neutral Higgs bosons ($\fH$) in
the $s$ channel in sufficient quantity to study their properties
directly (remember that $m_{\mu}\simeq 200\, m_e$). The crucial point
is also that in spite of the fact that the $\mu^+\mu^-\fH$ coupling,
being proportional to $m_\mu$, is still small, if the Muon Collider is
run on the Higgs resonance, $\sqrt{s}=m_\fH$, Higgs bosons may be
produced at an appreciable rate.

\begin{figure}[htb]
\leavevmode
\begin{center}
\resizebox{!}{6cm}{%
\includegraphics{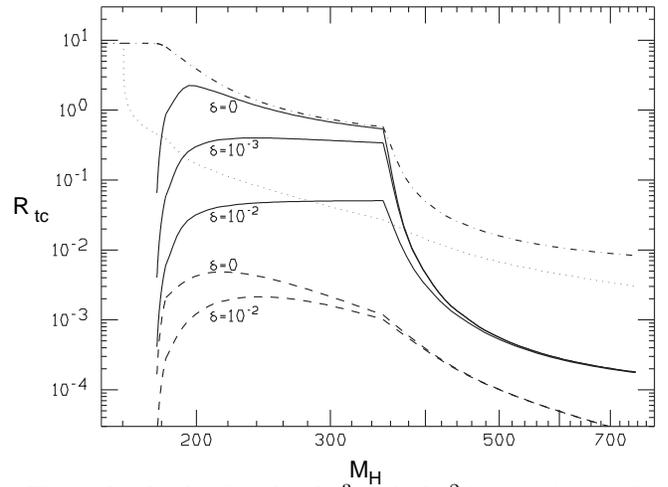}}
\caption[]{$R_{tc}$ for $\delta=0$,
$10^{-3}$ and $10^{-2}$ in case~1 (set of solid curves) and case~2
(set of dashed curves). We also plot $\tilde R(\fH)$ in case~1
(dot-dashed) and case~2 (dotted).}
\label{mumu_all}
\end{center}
\end{figure}

We have considered \cite{mumutc} the simple but fascinating
possibility that such a Higgs, $\fH$, has a flavor-changing $\fH t\bar
c$ coupling, as is the case in Model III or in any other 2HDM with
FCNC. As we did for the $e^+e^-$ case, also in the $\mu^+\mu^-$ case
we can define the analogous of $R_{tc}$ in eq.(\ref{rtc_ee}) to be

\begin{equation}
R_{tc}= \tilde R(\fH)\,(B^\fH_{t\bar c}+B^\fH_{c\bar t})
\label{rtc_mumu}
\end{equation}

\noindent where $\tilde R(\fH)$ is the effective rate of Higgs
production at a Muon Collider with beam energy spread described by
$\delta$ (i.e., $m_\fH^2(1-\delta)<s<m_\fH^2(1+\delta)$)

\be
\tilde R(\fH)=\left[ \frac{\Gamma_\fH}{m_\fH\delta} \arctan
\frac{m_\fH\delta}{\Gamma_\fH} \right] R(\fH) \label{rtildef}
\ee

\noindent $R(\fH)$ is here the rate of Higgs production, $\Gamma_\fH$
the width of the considered Higgs and $B^\fH_{t\bar c}$ or
$B^\fH_{c\bar t}$ denotes the branching ratio for $\fH\rightarrow
t\bar c$ and $\fH\rightarrow c\bar t$ respectively.  Assuming that the
background will be under reasonable control by the time they will
start operate a Muon Collider, our extimate is that $10^{-3}<\tilde
R_{tc}\le 1$, depending on possible different choices of the
parameters. In Fig.~\ref{mumu_all} we have illustrated in particular
the case in which $\fH\!=\!h^0$, and $\alpha\!=\!0$ (case 1) or
$\alpha\!=\!\pi/4$ (case 2).  We extimate that for a Higgs particle of
$m_\fH=300$ GeV, a luminosity of $10^{34} cm^{-2} s^{-1}$ and a year
of $10^7 s$ ($1/3$ efficiency), a sample of $tc$ events ranging from
almost one hundred to few thousands can be produced
\cite{mumutc}. Given the distinctive nature of the final state and the
lack of a Standard Model background, the predicted luminosity should
allow the observation of such events. Therefore many properties of the
Higgs-tc coupling could be studied in detail.
\vspace{.5cm}

Finally we want to consider the impact that a tree level $\xi^U_{ct}$
coupling could have on the present scenario of the Higgs discovery.
As was already pointed out in the literature \cite{hou}, if
$M_\fH>m_t$ (for $\fH=\bar H^0,h^0$ or $A^0$) Model III allows the new
decay channel

\be
\fH\rightarrow c\bar t +\bar c t
\ee

\noindent which should also be considered in the search for a non
standard Higgs particle. In the mass range $m_t<M_{\fH}<2m_t$, this
single top production is of particular interest because its rate can
be greater than the rate for $\fH\rightarrow b\bar b$ while the decay
$\fH\rightarrow t\bar t$ is not yet possible. Assuming
Eq.~(\ref{coupl_sher}), the rate for $\fH\rightarrow c\bar t +\bar c
t$ is given by

\bea
\label{rate_hct}
\Gamma(\fH\rightarrow c\bar t +\bar c t)&=&
N_c\frac{G_F}{4\sqrt{2}\pi}M_\fH\lambda_{ct}^2 m_c m_t\cdot\\
&&\!\!\!\!\!\!\!\!\!\!\!\!\!\!\!\!\!\!\!\!\!\!\!\!\!\!\!\!\!
\left[1-\frac{(m_t+m_c)^2}{M_\fH^2}\right]^{3/2}
\left[1-\frac{(m_t-m_c)^2}{M_\fH^2}\right]^{1/2}\nonumber
\eea

\noindent to be compared with the rate for $\fH\rightarrow q\bar q$,
i.e.

\be
\Gamma(\fH\rightarrow q\bar q)= N_c\frac{G_F}{4\sqrt{2}\pi}
M_\fH m_q^2\left[1-4\frac{m_q^2}{M_\fH^2}\right]^{3/2}
\label{rate_Hqq}
\ee

\noindent We see for instance that for $M_\fH\simeq 300$ GeV,
$\Gamma(\fH\rightarrow c\bar t +\bar c t)\sim 6 \lambda_{ct}^2
\Gamma(\fH\rightarrow b\bar b)$. 
Therefore, depending on $\lambda_{ct}$, there are cases in which in
the range $m_t<M_{\fH}<2m_t$ we could predict a distinctive signal,
both with rispect to the SM and to SUSY. When $\fH\!=\!h^0,\bar H^0$
then $\fH\rightarrow c\bar t+t\bar c$ competes only with the decays
$\fH\rightarrow ZZ \,\,\mbox{or}\,\, WW$, depending on the value of
the phase $\alpha$. In the case $\fH\!=\!A^0$ the decays into gauge
boson pairs are absent.

When the phase $\alpha$ is chosen in such a way that the couplings
$\fH ZZ$ and $\fH WW$ (for $\fH=h^0,\bar H^0$) are suppressed, the
decay we are interested in can be produced for instance via
$e^+e^-\rightarrow h^0A^0,h^0\bar H^0\rightarrow (t\bar c+c\bar t)
\,q\bar q$ (lepton collider) or $gg\rightarrow \fH\rightarrow t\bar
c+c\bar t$ (hadron collider).  Therefore both NLC and LHC should be
able to look for it: the first one would offer the possibility of a
much cleaner signal while the second one would provide a much higher
statistics. As is the case of many other decays, a good b-tagging is
clearly necessary. However the kinematic constraints of the
$\fH\rightarrow t\bar c +c\bar t$ decay should be so distinctive to
limit the size of the background.  We think that dedicated simulations
and sistematic studies of the background will be useful in
understanding the real potentiality of this decay channel.

\vspace{0.5cm}
In conclusion, we think that Model III offers a simple but interesting
example in which some important topics of the physics at the future
colliders can be investigated. With a few assumptions we are able to
propose some distinctive processes, the existence of which would be
clear evidence of some very new physics beyond the Standard Model.

\end{document}